\documentclass[iop]{emulateapj}
\usepackage{amsmath,color}
\usepackage{array}
\usepackage{float}
\usepackage{graphicx}
\usepackage{subfigure}

\usepackage{natbib}

\begin{document}
\title{A hypothesis for the color bimodality of Jupiter Trojans}
\author{Ian Wong and Michael E. Brown}
\affil{Division of Geological and Planetary Sciences, California Institute of Technology,
Pasadena, CA 91125, USA; iwong@caltech.edu}

\begin{abstract}
One of the most enigmatic and hitherto unexplained properties of Jupiter Trojans is their bimodal color distribution. This bimodality is indicative of two sub-populations within the Trojans, which have distinct size distributions. In this paper, we present a simple, plausible hypothesis for the origin and evolution of the two Trojan color sub-populations. In the framework of dynamical instability models of early Solar System evolution, which suggest a common primordial progenitor population for both Trojans and Kuiper belt objects, we use observational constraints to assert that the color bimodalities evident in both minor body populations developed within the primordial population prior to the onset of instability. We show that, beginning with an initial composition of rock and ices, location-dependent volatile loss through sublimation in this primordial population could have led to sharp changes in the surface composition with heliocentric distance. We propose that the depletion or retention of H$_{2}$S ice on the surface of these objects was the key factor in creating an initial color bimodality. Objects that retained H$_{2}$S on their surfaces developed characteristically redder colors upon irradiation than those that did not. After the bodies from the primordial population were scattered and emplaced into their current positions, they preserved this primordial color bimodality to the present day. We explore predictions of the volatile loss model --- in particular, the effect of collisions within the Trojan population on the size distributions of the two sub-populations --- and propose further experimental and observational tests of our hypothesis

\end{abstract}

\section{Introduction}
In the past decade, Jupiter Trojans have been the subject of increasing scientific interest, and understanding their origin and surface properties promises to unlock many of the fundamental aspects of Solar System formation and evolution. These minor bodies share Jupiter's orbit around the Sun at 5.2~AU and reside in two clusters at the stable L4 and L5 Lagrangian points. One of the most important discoveries about the physical properties of Trojans is the existence of two sub-populations --- the less-red (LR) and red (R) Trojans --- whose members differ categorically with respect to several photometric and spectroscopic properties. Bimodality has been reported in visible and near-infrared colors \citep{szabo,roig,emery}, as well as in longer wavelength reflectance \citep{grav}. The two sub-populations are present in both the L4 and L5 swarms. Meanwhile, members of the only robustly attested collisional family within the Trojans --- the Eurybates family at L4 \citep{broz} --- all belong to the less-red sub-population \citep{fornasier}, an observation that has important implications for the interplay between self-collisions and the surface properties of these objects.

While recent spectroscopic study in the near-infrared has detected water ice on the surfaces of some large Trojans, along with weaker absorption features possibly attributable to organics \citep{brown2015}, the overall surface compositions of Trojans remain poorly constrained. In particular, the color bimodality has no current explanation. Although some theories have been forwarded to produce the range of Trojan colors through ongoing processes such as surface gardening and/or various irradiation mechanisms \citep[e.g.,][]{melita}, none of these theories predict a bimodality in color, instead yielding a single broad color distribution that is inconsistent with observations. On the basis of the differing spectral properties of LR and R Trojans,\citet{emery} suggested that perhaps one of the two sub-populations formed in the Main Belt asteroid region, while the other formed in the outer Solar System before being captured by Jupiter.

Recent theories of Solar System evolution, however, suggest that after an early dynamical instability between giant planets, the Jupiter Trojan region was filled exclusively with objects that formed at large heliocentric distances within an extended planetesimal disk situated beyond the primordial orbits of the ice giants \citep{morbidelli,roig2}. In the context of these dynamical instability models, both LR and R Trojans must have originated from the same primordial population. To date, no hypothesis has been proposed to explain the existence of two distinct sub-populations from a single source region.

Support for the idea of a shared formation environment for the Trojans and Kuiper belt objects (KBOs) comes from the observation that the size distributions of Trojans and hot (i.e., dynamically excited) KBOs are consistent with each other \citep{fraser2014}. In addition, the total Trojan size distribution for objects smaller than $\sim$100~km is consistent with collisional equilibrium \citep[e.g.,][]{marzari}. Unexpectedly, however, the size distributions of the individual LR and R Trojan sub-populations are highly distinct; specifically, the power-law slopes of the magnitude distributions for objects smaller than $\sim$100~km in diameter are discrepant, revealing a monotonic increase in the relative number of LR Trojans with decreasing size \citep{wong,wong2}. 

Such distinct size distributions are difficult to reconcile with the hypothesis that the Trojans share a single source population. However, \citet{wong} demonstrated that, starting from an initial state where both sub-populations had identical magnitude distributions, a process wherein the collisional fragments of both R and LR Trojans become LR objects would naturally account for the relative depletion of R Trojans and the simultaneous enrichment of LR Trojans with decreasing size, as well as the collisional equilibrium of the overall population. While this explanation did not appeal to specific chemical or physical processes for the color conversion, the consistency between simulated and observed magnitude distributions suggests that R-to-LR conversion may provide a promising angle from which to address the question of the Trojans' color bimodality and the different LR and R size distributions.

In this paper, we propose a simple, chemically plausible hypothesis, developed within the framework of current dynamical instability models, to explain the origin of the color bimodality within the Trojans and the differing size distributions of the two color sub-populations. Central to our hypothesis is location-dependent volatile loss, which established sharp changes in \textit{surface} composition and, thus, color across the primordial trans-Neptunian planetesimal disk prior to the onset of dynamical instability. In particular, we suggest that the loss or retention of H$_{2}$S due to sublimation resulted in two types of surface chemistry among objects in the primordial disk, subsequently leading to different characteristic colors upon irradiation. In Section~\ref{sec:obs}, we briefly summarize the present understanding of both Trojan and KBO colors, which serves to motivate and constrain our hypothesis. The  location-dependent volatile loss model is detailed in Section~\ref{sec:volatiles}, and finally, the predictions of our hypothesis with respect to surface colors, collisions, and size distributions are discussed in Section~\ref{sec:colors}.

\section{Colors of Trojans and KBOs}\label{sec:obs}
Recent dynamical instability models of the early Solar System posit a single primordial body of planetesimals from which the current Jupiter Trojans and KBOs are sourced. Therefore, the observed colors of both Trojans and KBOs are important in forming the foundational premise for the hypothesis developed in this paper, as well as establishing the range of phenomena that the hypothesis strives to explain.

\begin{figure}[t!]
\begin{center}
\includegraphics[width=9.25cm]{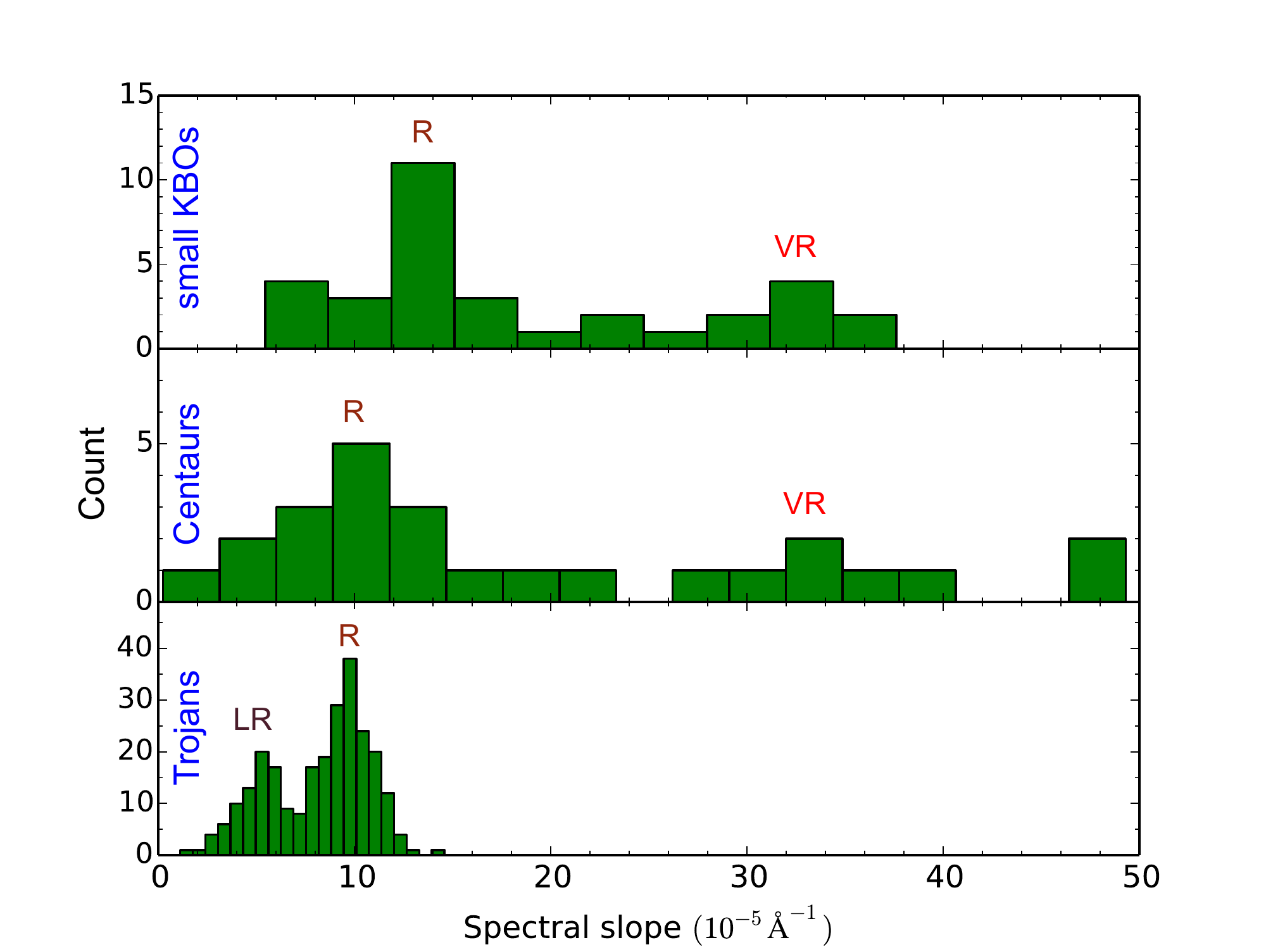}
\end{center}
\caption{Histograms of the measured spectral slope values of small KBOs (top panel), Centaurs (middle panel), and Trojans (bottom panel). The small KBO and Centaur color distributions include only objects fainter than $H=7$ and spectral slope uncertainties smaller than the bin size. Cold classical KBOs have been filtered out by omitting all KBOs with inclinations less than $5^{\circ}$. Color bimodalities are evident in all three minor body populations; they are labeled with their corresponding relative colors: less-red (LR), red (R), and very red (VR).} \label{colors}
\end{figure}

Analyses of the visible spectral slopes of Trojans derived from Sloan Digital Sky Survey photometry reveal a color distribution with two peaks, corresponding to the LR and R sub-populations \citep{roig,wong}. The spectral slope distribution of Trojans brighter than $H=12.3$ is shown in the bottom panel of Figure~\ref{colors}; the LR sub-population has colors centered around $5\times 10^{-5}~\AA^{-1}$, while the R sub-population has a mean color of $\sim$$10\times 10^{-5}~\AA^{-1}$. Aside from their differences in spectral properties, both LR and R Trojans have notably uniform low albedos averaging around 4\% \citep[e.g.,][]{grav}. 

The color distribution of hot KBOs in the same size range as Trojans ($7.0< H < 12.3$), compiled from spectral slopes listed in the MBOSS database \citep{hainaut}, is shown in the top panel of Figure~\ref{colors}. In the middle panel, we also present the catalogued spectral slope data for similarly-sized Centaurs, which are former KBOs that have been scattered onto short-lived giant planet-crossing orbits. Many color measurements of KBOs and Centaurs in the literature have large uncertainties, and we have chosen to only include those measurements with uncertainties smaller than the bin size ($\sim$$3\times 10^{-5}~\AA^{-1}$). In addition, we have omitted objects with inclinations less then $5^{\circ}$ in order to remove the bulk of the cold classical KBOs. The cold classical KBOs are a population of low-inclination objects beyond 41~AU with physical properties that are markedly distinct from those observed across the rest of the Kuiper belt \citep[see, for example, the review by][]{brown2012} and may have experienced a different formation and evolutionary history than the rest of the KBO population \citep[e.g.,][]{batygin}. Given the possibility that the cold classical KBOs do not share the same progenitor population as the remainder of the KBOs and the Trojans, we do not consider them in the present work.

The color distributions of both KBOs and Centaurs have a notable peak at around $10\times 10^{-5}~\AA^{-1}$, with spectral slopes extending to much higher values. An overabundance of spectral slopes higher than $30\times 10^{-5}~\AA^{-1}$ is discernible, suggesting an underlying bimodality in the KBO and Centaur color distributions. The bimodal nature of the  Centaur color distributions has been thoroughly discussed in the literature \citep[e.g.,][]{peixinho2,barucci3,perna}, and recent work has suggested a similar bimodality for small KBOs, i.e., KBOs in the same size range as that of the known Centaurs \citep[e.g.,][]{fraserbrown,peixinho,lacerda}. It is therefore plausible that both (non cold classical) small KBOs and Centaurs are comprised of two color sub-populations, hereafter referred to as the red (R) and very red (VR) KBOs and Centaurs. The color terminology has been chosen to reflect the relative colors of the various sub-populations, as shown in Figure~\ref{colors}. One important difference between the KBOs/Centaurs and Trojans is the albedo distribution. Both R KBOs and R Centaurs have very low albedos that are comparable with the Trojans (averaging around $4-6\%$), while VR KBOs and Centaurs are noticeably brighter, with mean albedos in range $11-12\%$ \citep[e.g.,][]{stansberry,fraser2014}. 

Importantly, no correlation between color and any dynamical or orbital parameter has been found in either the Trojans \citep{fornasier,melita2008,emery} or the KBOs \citep[e.g.,][]{morbidellibrown,doressoundiram}. This indicates that the observed color bimodalities cannot be attributed to the local conditions at different locations within the Trojan and Kuiper belt regions.  The lack of correspondence between color and present-day environment points toward the color bimodality being primordial (i.e., already present when the Trojans and KBOs were emplaced in their current locations).

\section{Volatile loss model}\label{sec:volatiles}

Within the dynamical instability model, the observation that Trojans, Centaurs, and small KBOs all appear to have a bimodal color distribution that is not attributable to any aspect of the populations' current orbital/dynamical architecture suggests that the color bimodality developed early on within the primordial trans-Neptunian planetesimal disk. Large solid bodies that formed in this region moved through the disk as they accreted mass, thereby incorporating materials from diverse regions \citep[e.g.,][]{kenyon}. As a result, chemical gradients that existed in the disk initially were averaged over in the macroscopic bodies, whose bulk compositions would then reflect a mix of materials found throughout the disk. In light of this, a possible explanation for a color bimodality within the primordial disk (as opposed to a smooth gradient) is a physical process that results in sharp divisions in the surface chemistry of objects, regardless of any variations in the bulk composition across the primordial planetesimal disk.

In our hypothesis, we follow previous work on the surface colors of KBOs \citep{brown} by positing that the sublimation of volatile ices can yield sharp changes in the surface chemistry of objects as a function of formation location. Objects that formed in the primordial planetesimal disk were accreted out of a mix of rocky material and ice, including many volatile ices. We take this as the initial composition of objects in our model. Objects closer in to the Sun had higher surface temperatures and correspondingly higher volatile loss rates, and after some time, their surfaces would become depleted in the more volatile species. Meanwhile, objects that formed farther away would have retained the more volatile molecules on their surfaces. 

We update the results of \citet{brown} in two important ways. First, the objects that we consider here have diameters less than 250~km; this limit roughly corresponds to the size of the largest Trojan (624 Hektor). Due to their relatively small size, these objects would not have sustained a bound atmosphere that could act as a buffer against volatile loss. Quantitatively, this means that the value of the Jeans parameter (i.e., ratio of the escape velocity to the thermal energy of a gas molecule in an atmosphere) is much less than unity for all Trojans, assuming any plausible atmospheric composition and across all temperatures in the middle and outer Solar System. Second, we correct numerical errors in the \citet{brown} vapor pressures. We quantitatively model volatile loss via direct sublimation from a surface into vacuum. For a volatile ice species $i$ with molecular mass $m_{i}$, the rate of mass loss per unit surface area is directly related to the vapor pressure $p_{i}$ by the relation \citep[e.g.,][]{wallis}
\begin{equation}\dot{m}_{i}=p_{i}\sqrt{\frac{m_{i}}{2\pi k T}},\end{equation}
where $k$ is the Boltzmann constant, and $T$ is the blackbody temperature of the surface in equilibrium with incident solar irradiation, assuming reradiation across the entire surface: $T\equiv \left(S_{0}(1-A)/4 \sigma a^{2}\right)^{1/4}$. Here, $\sigma$ is the Stefan-Boltzmann constant, $A$ is the albedo, $a$ is the heliocentric distance in AU, and $S_{0}=958$~W/m$^2$ is the solar irradiance at 1~AU, assumed at this early stage of Solar System evolution to be 70\% of its current value \citep{gough}. We use vapor pressures for each ice species as computed by \citet{fray}.

While the typical albedo of these objects after surface radiation processing is very low, as reflected by present-day observations, it is expected that these icy bodies were initially brighter. With a surface composition comprised of primarily water ice and rocky material (see below), the initial albedo would likely have exceeded 0.5. Through irradiation, the surface was continuously altered, reddening and darkening as described in Section~\ref{subsec:rad}. Meanwhile, the active collisional environment of the early Solar System produced a steady gardening of the surface, exposing brighter pristine and unirradiated material from the interior. Therefore, in our modeling, we chose arbitrary intermediate values between the current albedo and the expected initial albedo. For the results shown in this paper, we have assumed $A=0.2$; changes to the assumed albedo by factors of $2-3$ in either direction do not substantially affect the general conclusions of the model. 

The time $t_{i}$ needed to deplete a surface layer with thickness $x$ of a particular volatile ice species is given by
\begin{equation}t_{i}=\frac{\rho\eta\beta_{i} x}{\dot{m}_{i}},\end{equation}
where $\rho$ is the bulk density of the object, $\eta$ is the fraction of the object's mass that is ice, and $\beta_{i}$ is the abundance of the volatile ice $i$ relative to water ice. The rate of sublimation is a steep exponential function of temperature, and therefore, most model parameters have little impact on the final result. 

Spectroscopic studies of Trojans have revealed features indicative of both water ice \citep{brown2015} and fine-grained silicates \citep{emeryold} on the surface. For the calculations presented here, we have assumed $\rho=1.0$~g/cm$^{3}$, which is comparable to the measured Trojan densities in the literature \citep[e.g., $0.8^{+0.2}_{-0.1}$~g/cm$^{3}$ for 617 Patroclus;][]{marchis}. This low density suggests a significant ice fraction, as well as high porosity. In this paper, we present results for the case of $\eta=0.5$ to represent an ice-to-rock ratio of unity. We note that using $\eta=1$ instead, corresponding to a 100\% ice fraction, does not have a noticeable impact on the results. 

Objects that formed in the primordial trans-Neptunian disk would have accreted with a roughly cometary composition, with rocky material and a wide variety of volatile ices. Published spectra of Trojans have hitherto not revealed any incontrovertible features except for some water ice absorption; meanwhile, obtaining high quality spectra of similarly sized Centarus and KBOs is difficult. Instead, we can use the detailed measured ice compositions of short-period comets as a proxy for the volatile component of these primordial planetesimals from the outer Solar System. In our model, we consider all volatile ice species with average measured abundances relative to water greater than 0.5\% as reported from comet comae measurements \citep{bockelee} --- CH$_{3}$OH, NH$_{3}$, CO$_{2}$, H$_{2}$S, C$_{2}$H$_{6}$, CO, and CH$_{4}$. We set all initial relative volatile ice abundances to $\beta_{i}=0.01$. We note in passing that many of these volatile ices, or the irradiation products thereof, have been detected on some of the larger KBOs \citep[see][and references therein]{brown2012,baruccibook}, so they are expected to have accreted onto the smaller primordial bodies in this region as well.

The thickness parameter $x$ must reflect the amount of depletion needed in order to effect a change in the surface chemistry and its response to irradiation. While the final color of an object depends only on the chemistry of the topmost layers, in an early Solar System environment characterized by frequent collisions, gardening and minor impacts constantly excavated the surface, exposing interior material to irradiation. Meanwhile, even kilometer-sized short-period comets have measurable amounts of highly volatile ices such as CO, indicating that volatile ices in the interior of these objects (deeper than $\sim$1~km) are protected from sublimation. 

The thickness of the surface layer that must be depleted is equal to the depth to which material is excavated through collisional activity. We estimate this by extrapolating the results from collisional modeling of the current Kuiper Belt to the expected conditions within the primordial trans-Neptunian disk. \citet{durda} computed collision rates and the effect of minor impacts on the surface of present-day KBOs and demonstrated that the cumulative fraction of the surface area of 2 to 200~km diameter objects cratered by impactors of radius greater than 4~m ranges from a few to a few tens of percent over $\sim$4 Gyr. For reasonable assumptions of the surface material properties, they show that such impacts are expected to excavate down to a depth of at least $40-50$~m.

The number of objects in the primordial trans-Neptunian region prior to the onset of dynamical instability is expected to have exceeded the current KBO population by roughly two orders of magnitude \citep[e.g.,][]{tsiganis}. Therefore, we expect that, in the period before the disruption of the primordial trans-Neptunian planetesimal disk (see below), all objects in the observable size range for Trojans would have had their entire surface reworked through collisions. Many of the cratering events would have been caused by impactors significantly larger than tens of meters in diameter, so a significant fraction of the impacts would have exposed material to a depth of many hundreds of meters or more. In this paper, we have taken $x=100$~m to be the thickness of the surface layer that is affected by these impacts.

\begin{figure}[t!]
\begin{center}
\includegraphics[width=9.25cm]{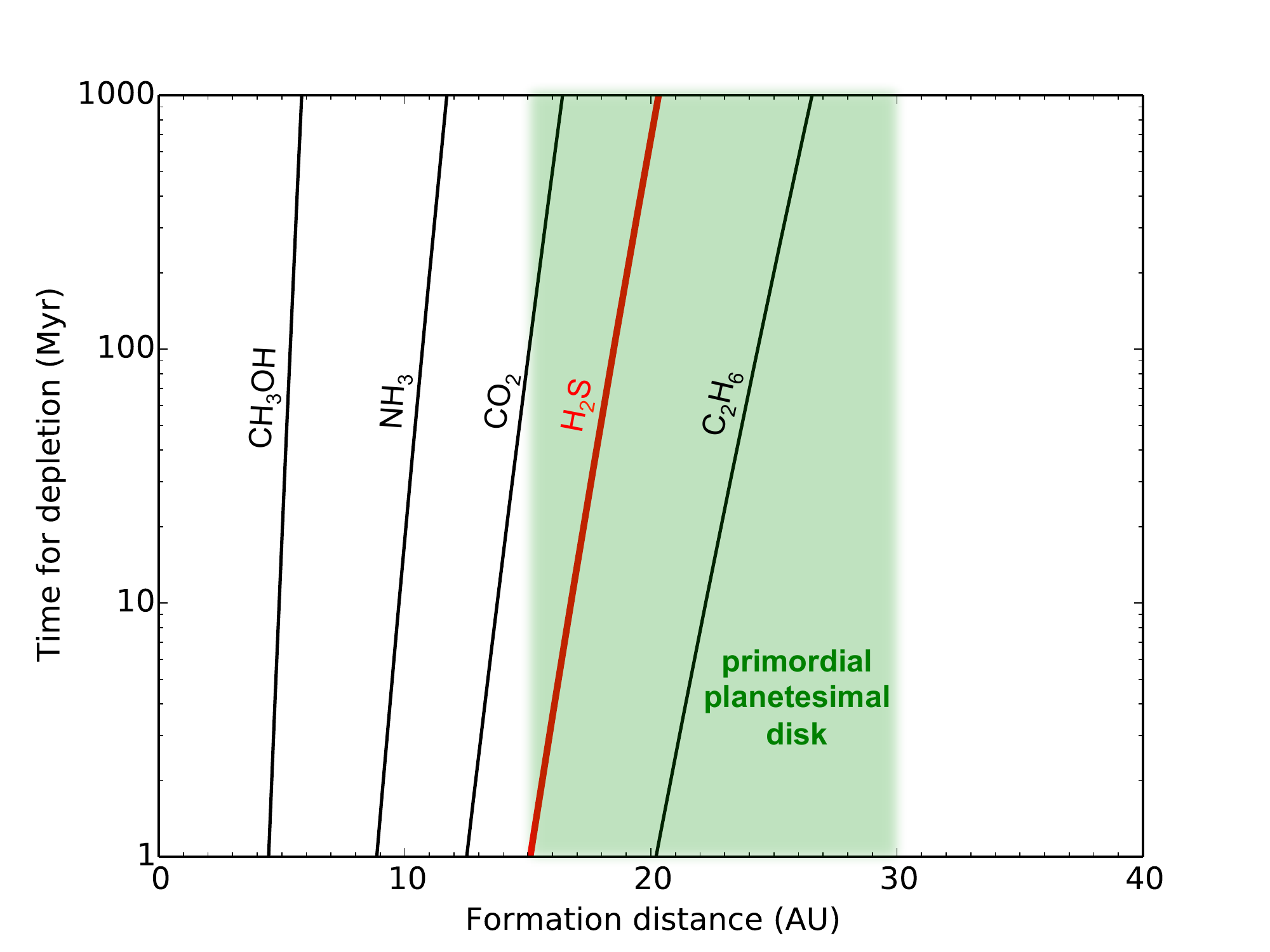}
\end{center}
\caption{Sublimation lines in the early Solar System, as a function of formation distance. Objects located to the left of a line have the respective volatile ice species depleted from a 100~m layer at the surface after the corresponding elapsed time on the y-axis, while objects located to the right of a line have retained the respective volatile ice species on the surface. The approximate location of the pre-instability primordial planetesimal disk is depicted by the green band. Notably, the H$_{2}$S sublimation line (in red) passes through the region of the primordial disk.} \label{massloss}
\end{figure}

The relevant timescale for our model is the irradiation timescale, i.e., the time required for incident radiation to change in the properties of the surficial material. Experiments on volatile ices have demonstrated that the effects of irradiation depend primarily on the total energy deposited and saturate after a certain energy threshold is reached. In the context of surface colors, only the top $\sim$1~$\mu$m of the surface material is sampled by visible and near-infrared observations. \citet{hudson2008} estimated that the radiation dose at $\sim$1~$\mu$m depth for objects at 20~AU (roughly the center of the primordial disk in canonical simulations of dynamical instability models) is $1000-10000$~eV/16~amu over the lifetime of the Solar System. Using the saturation doses for volatile ice irradiation reported by \citet{brunetto} ($100-300$~eV/16~amu), we arrive at estimates of the irradiation timescale between 50~Myr and 1~Gyr.  In this paper, we set the irradiation timescale at 100~Myr; values differing by an order of magnitude in either direction do not yield qualitatively different conclusions.

The canonical version of current dynamical instability models aligns the onset of instability  with the Late Heavy Bombardment, which occurred $\sim$650~Myr after the formation of the Solar System \citep{gomes}. Comparing this estimated disk lifetime to our choice of irradiation timescale, we find that irradiation processing is expected to have proceeded to saturation prior to the scattering of the primordial disk.

Figure~\ref{massloss} shows the time needed to deplete the top 100~m of an object of various volatile ice species, as a function of heliocentric distance. The curves denote sublimation lines: objects that formed closer than a line would have surfaces depleted of the corresponding volatile ice species, while objects that formed farther would have retained it. Simulations of early Solar System evolution predict an initially compact planetary configuration, with a planetesimal disk beyond the orbit of the outermost ice giant extending from 15~AU to about 30~AU, indicated in the figure by the green band \citep{tsiganis, levison2008}. CH$_{3}$OH, NH$_{3}$, and to a certain extent CO$_{2}$ are involatile throughout the entire primordial disk. (We note that the suggestion in \citet{brown} that CH$_{3}$OH had a sublimation line within the primordial planetesimal disk was mainly a result of a numerical error in their vapor pressures.) Objects in the inner part of the disk have surfaces depleted in H$_{2}$S after an elapsed time comparable to the irradiation timescale of 100~Myr, while objects in the outer part retain H$_{2}$S. The farthest most objects also retain C$_{2}$H$_{6}$. The sublimation lines for CO and CH$_{4}$ are located at much larger heliocentric distances, out of the range shown in Figure~\ref{massloss}.

\section{Surface colors}\label{sec:colors}
For a wide range of assumptions, volatile loss in the primordial planetesimal disk would have occurred as shown in Figure~\ref{massloss}. Given the sublimation-driven surface chemistry developed by the model in the previous section, we now consider the effects of radiation processing on the surface properties and examine the predictions of our hypothesis with respect to the surface colors of Trojans and KBOs.

\subsection{Surface radiation processing}\label{subsec:rad}

Objects throughout the primordial disk retained methanol, ammonia, and carbon dioxide ices on their surfaces. While no laboratory experiments have studied the effects of irradiation on analogous surfaces, \citet{brunetto} demonstrated that the irradiation of single-component methanol ice yields a darkened and moderately-reddened irradiation crust comprised of organic residue, with spectral slopes in the visible through the near-infrared comparable to values measured for many R KBOs and Centaurs.  We therefore propose that the presence of methanol ice on the surface is the primary factor that led to surface colors characteristic of R objects upon irradiation. It is interesting to note that methanol is the only involatile molecule other than water that has been identified in the spectrum of either a centaur \citep{cruikshank} or a KBO \citep{barucci,barucci2,brown2012}.

Objects farther out in the primordial disk additionally retained H$_{2}$S on their surfaces. Irradiation studies of sulfur-bearing ices have found that among the stable irradiation products are long-chained sulfur molecules \citep[e.g.,][]{moore1984,cassidy}, which produce a dark residue and have been interpreted, for example, as the cause of the darkened, intensely red deposits found on Io \citep{carlson}. Within the framework of our hypothesis, we propose that the retention of H$_{2}$S on the surface of objects in the outer primordial disk contributed to a significant additional reddening, leading to the VR colors. The only remaining sublimation line in the region of the primordial planetesimal disk is C$_{2}$H$_{6}$. Experiments on mixtures of water ice and ethane ice yield similar irradiation products as mixtures with methanol or carbon dioxide ice \citep[e.g., complex hydrocarbons][]{moore1998,hudson2009}, so it is not expected that the addition of ethane ice on the surface would entail any notable alterations in the surface color.

\begin{figure*}[t!]
\begin{center}
\includegraphics[width=15cm]{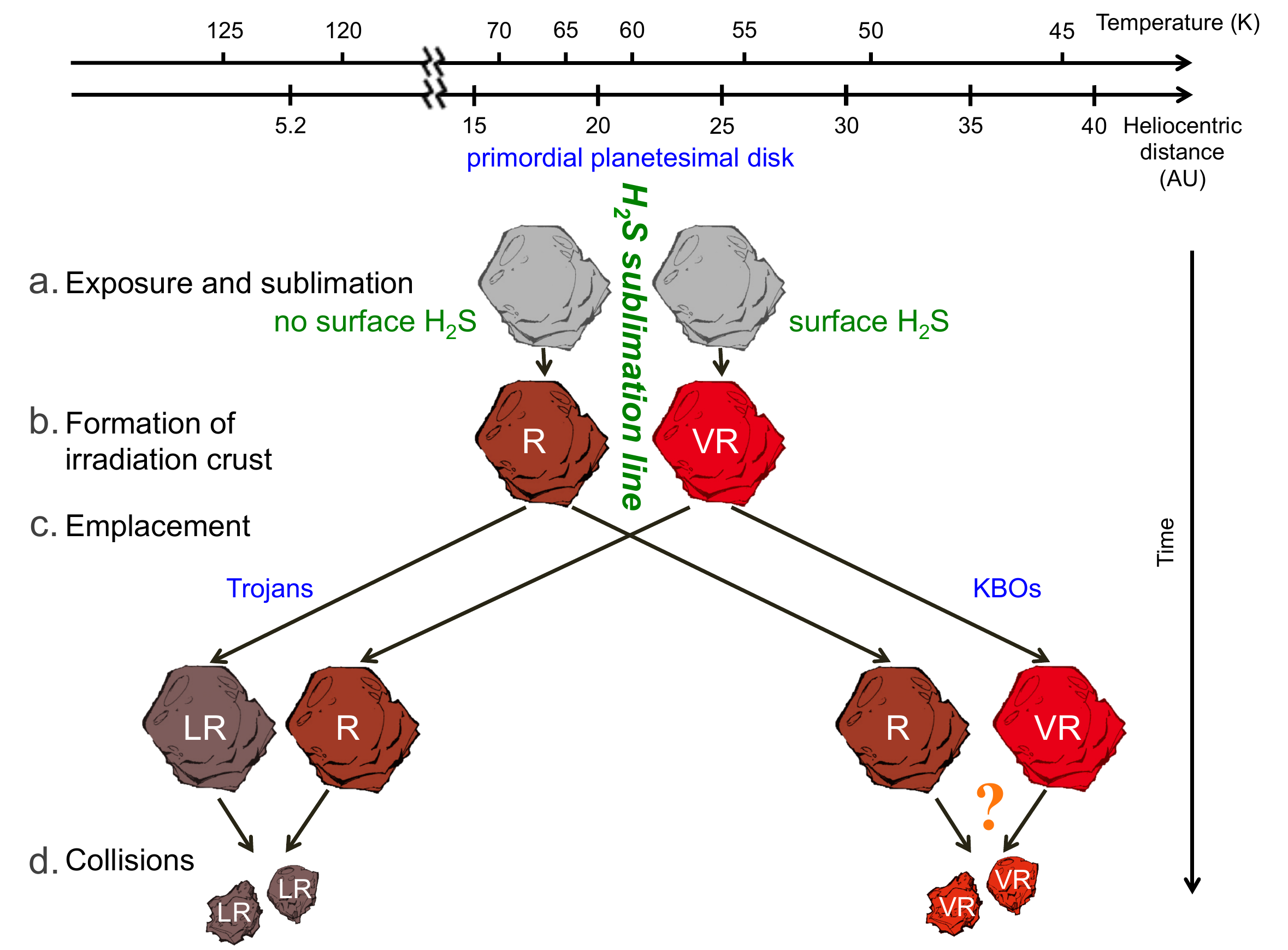}
\end{center}
\caption{Pictoral representation of the hypothesis presented in this paper and the major predictions thereof. Panel a depicts the initial state of bodies within a primordial planetesimal disk extending from 15~AU to roughly 30~AU at the moment when the nebular gas dispersed and the objects were first exposed to sunlight. Each body was comprised of a pristine mix of rocky material and ices; volatile ice sublimation began. Panel b illustrates the development of the primordial R and VR colors: Objects that formed inside of the H$_{2}$S sublimation line were depleted in H$_{2}$S after $\sim$100~Myr and formed a dark, reddish irradiation crust (due primarily to retained methanol). Meanwhile, objects that formed farther out beyond the H$_{2}$S sublimation line retained H$_{2}$S on their surfaces and thus developed a much redder coloration. The emplacement of Trojans and KBOs during the Nice Model dynamical instability is denoted in panel c. We expect that the primordial R and VR objects became R and VR KBOs and Centaurs, respectively, while the Trojans experienced surface color evolution, maintaining the initial color bimodality but resulting in relatively less red colors. Panel d describes the result of collisions in the current Trojan and KBO populations: The surfaces of Trojan collisional fragments of Trojans become depleted in all volatile ice species, thereby becoming LR objects. Although no observational evidence exists at present, we propose that the collisional fragments of small KBOs likely retain H$_{2}$S on their surfaces, eventually developing into VR objects.} \label{diagram}
\end{figure*}

We have forwarded a speculative explanation for the development of a primordial color bimodality within the primordial planetesimal disk, with the depletion vs. retention of H$_{2}$S ice on the surface as the main distinguishing factor between R and VR objects, respectively.  This process is outlined in panels a and b of Figure~\ref{diagram}. We note that while the overall darkening of primordial disk objects upon irradiation is predicted by the hypothesis, the different albedos of R and VR KBOs and Centaurs (see Section~\ref{sec:obs}) remains unexplained here, since experimental data of analogous surfaces do not exist so as to allow any comparison between cases with and without H$_{2}$S.

\subsection{Emplacement}

The next step in the evolution of the primordial disk is the period of dynamical instability, during which a portion of the objects was scattered outward to populate the present-day Kuiper belt. Given that the color alteration caused by irradiation is likely to have proceeded to saturation within the primordial disk (see Section~\ref{sec:volatiles}), we expect that the colors of the irradiation mantles on KBOs have remained unchanged in the roughly 4~Gyr that has elapsed since these objects were emplaced in their current orbits beyond 30~AU.  The average surface temperature on KBOs that scatter inward to become Centaurs do not markedly increase beyond those that were experienced early on in the primordial disk, so their surfaces are not expected to evolve significantly, thus providing an explanation of the statistically identical color distribution of KBOs and Centaurs (Section~\ref{sec:obs}). 

According to dynamical instability models, a portion of the primordial disk was scattered inward and later captured by Jupiter to become the present-day Trojans \citep{morbidelli,roig2}. We recall from Figure~\ref{colors} that the LR and R Trojan color modes are situated at relatively more neutral colors than the two color modes in the KBOs and Centaurs. The conditions at 5.2~AU are characterized by higher surface temperatures and more intense irradiation than the conditions in the primordial planetesimal disk and may have altered the chemistry and physical properties of the irradiation mantle. Experiments simulating the effect of long-term irradiation on complex refractory organic residues (similar to the final irradiation products in methanol ice studies) have shown that at high radiation doses and temperatures, carbonization of the surface begins, which produces an overall neutralizing and darkening effect on the surface \citep[e.g.,][]{moroz}. This process could explain the alteration in surface colors of Trojans upon emplacement, predicting that the primordial R and VR objects become LR and R Trojans, respectively. Panel c in Figure~\ref{diagram} illustrates the emplacement process of both Trojans and KBOs.

\subsection{Collisions}

Lastly, we use our hypothesis to speculate on the result of mutual collisions within the Trojan and KBO populations since emplacement. When two Trojans are involved in a shattering collision, the thin irradiation crust on the surface of both objects is destroyed. The surface composition of the resultant fragments thus corresponds to the interior composition of Trojans, i.e., a mixture of water ice and rocky material, along with volatile ices that were shielded from sublimation at depth. At the current Trojan surface temperature of $120-125$~K, all the volatile ice species on the surface rapidly sublimate away, leaving behind only water ice and the rocky component. (Recall that the sublimation lines in Figure~\ref{massloss} are assuming 70\% of current solar flux. The present methanol sublimation line is located $\sim$3~AU farther from the Sun, well away from the Trojan region.) Without either CH$_{3}$OH or H$_{2}$S present on the surface, subsequent irradiation would not produce the characteristic R and VR colors that developed in the more distant primordial disk, and the resulting colors would be relatively less-red. Starting from an initially neutral surface, with likely higher albedo than the uncollided Trojans, space weathering processes would probably proceed in a similar fashion to scenarios reported for other middle Solar System bodies, e.g., darkening and reddening of the spectrum via nanophase iron on silicate grains \citep[e.g.,][]{bennett}.

The previous discussion provides the chemical basis for the R-to-LR conversion argument proposed in \citet{wong} and described in the Introduction. The collisional probability is inversely proportional to object size, which leads to an increase in the fraction of objects that are collisional fragments with decreasing size. All of the collisional fragments in the Trojan population are depleted in CH$_{3}$OH and H$_{2}$S, so regardless of whether the parent bodies are R and LR Trojans, the resulting fragments become LR objects. Therefore, the observed increase in LR-to-R number ratio with decreasing size is a natural outcome of the hypothesis outlined in this paper. The effect of collisions on Trojan colors is illustrated in panel d of Figure~\ref{diagram}.

An interesting prediction of our hypothesis is that the fragments from shattering collisions may constitute a third color sub-population of neutral objects. The only robustly attested collisional family in the Trojans, the Eurybates family, appears to be unique in having members that are exceptionally neutral \citep[e.g.,][]{fornasier}, in line with the prediction of our hypothesis. One might imagine that the eventual colors of Trojan collisional fragments, with surfaces completely depleted of volatile ices, may be distinct from the colors of uncollided LR Trojans, which arose from the evolution of primordial R (or possibly VR) objects emplaced during the dynamical instability. Further measurements of the colors and albedos of collisional fragments and smaller Trojans in general promise to provide a powerful observational test for the details of the R-to-LR conversion model.

The relatively low intrinsic collisional probability of Trojans means that shattering impacts on objects larger than a few tens of km in diameter are rare. More common are sub-catastrophic, cratering collisions, which garden the surface and can potentially excavate primordial, neutral-colored, unirradiated material from the interior. The collisional probability and impact velocities within the Trojan swarms are comparable to those found in the main belt \citep[e.g.,][]{marzari}, which in turn have been shown to yield a similar level of collisional activity as in the current Kuiper Belt \citep{durda}. Since the thickness of the surface layers that was altered by irradiation prior to emplacement is on the order of $x=100$~m, impacts would need to excavate through a depth greater than $x$ in order to expose pristine interior material. Using the results from \citet{durda}, we can expect that the cumulative surface fraction of Trojans that has been excavated past a depth of $\sim$100~m is at most a few tens of percent. We therefore conclude that the current level of collisional activity is insufficient to yield significant resurfacing and exposure of primordial material. The minor gardening within the Trojan population may contribute, as a secondary effect, to the dispersion of colors within each of the LR and R sub-populations.

At present, color measurements in the Kuiper belt have not yet probed to the sizes of collisional fragments. Nevertheless, within the framework of our hypothesis, it is possible to speculate on the consequences of mutual collisions between small planetesimals within the Kuiper belt. Upon shattering, the fragments from such a KBO collision would expose the interior composition of the objects --- water ice, rocky material, and all the volatile ice species that were present during the initial accretion. Unlike in the case of Trojans, the KBOs are currently in a colder environment. Crucially, throughout the entire Kuiper belt ($a>30$~AU), our volatile loss model predicts that H$_{2}$S is involatile. Therefore, all KBO collisional fragments would retain H$_{2}$S on their surfaces. We expect that after an initial state with pristine unirradiated ice on the surface, with likely higher albedo and a relatively neutral color, the total received photon and solar wind flux on the surfaces of these collisional fragments will turn them VR. In essence, these collisional fragments should evolve in an analogous way to the H$_{2}$S-retaining planetesimals in the primordial trans-Neptunian region, as discussed in Section~\ref{sec:volatiles}. We note that the fragments from the only known KBO collisional family to date --- the Haumea family --- came from the already volatile-depleted icy mantle of a large differentiated dwarf planet \citep{brownhaumea}, so they do not follow this proposed pattern.

\section{Conclusion}
We have proposed a simple hypothesis for the two color sub-populations within the Trojans. The main feature of our proposal is location-dependent sublimation of volatile ices within a primordial planetesimal disk in the outer Solar System, which divided objects in this region into two groups --- those that retained H$_{2}$S on their surfaces, and those that did not. As a result of irradiation chemistry on these initially pristine icy surfaces, these two groups developed different involatile surface colors, thereby imprinting an initial color bimodality into the population. The retention of H$_{2}$S on the surface acted as a significant reddening agent upon irradiation, leading to characteristically redder colors than in the face of H$_{2}$S-depleted surfaces. Following a period of dynamical instability, the objects in this primordial disk were scattered throughout the middle and outer Solar System, carrying the primordial color bimodality with them as they populated the Trojan region and the current Kuiper belt. Our hypothesis thus offers a chemically and dynamically plausible explanation for the color bimodality of both Trojans and KBOs, as summarized in Figure~\ref{diagram}. In addition, our model predicts that all collisional fragments in the Trojan swarms are compositionally identical and become relatively less-red upon irradiation, regardless of whether the parent body is a LR or a R object. Therefore, our hypothesis accounts for the relative depletion of R objects with decreasing size and offers a self-consistent explanation for the discrepancy between the observed LR and R Trojan size distributions.

Although it is supported by observational constraints and some experimental studies, the hypothesis advanced here is necessarily speculative. The most fruitful pathway for follow-up study would be systematic laboratory experiments that analyze the effects of irradiation on the color and reflectivity of surfaces analogous to the ones described in our hypothesis --- water ice-dominated mixtures of methanol, ammonia, and carbon dioxide ice, with or without the addition of H$_{2}$S ice. Crucially, these experiments must simulate the changing environments of Trojans and KBOs as they evolve within the primordial planetesimal disk and then are emplaced in their respective current locations. The main results that must be obtained in the experimental context for our hypothesis to be true are: (1) Irradiation of ice mixtures with methanol, ammonia, and carbon dioxide, but without H$_{2}$S, leads to reduction of albedo and a surface residue that has a color comparable to those seen on R KBOs and Centaurs. (2) Irradiation of ice mixtures with the addition of H$_{2}$S leads to reduction of albedo and a significantly redder irradiation mantle than in the absence of H$_{2}$S. (3) Intensified irradiation and higher temperatures alter the surface colors to those characteristic of Trojans. While some of these results are supported by laboratory data in the literature, only experimental verification using a uniform setup and conditions appropriate to the various minor body populations will allow us to confirm or refute the hypothesis for the color bimodality of Trojans presented here.

\acknowledgements
Discussions with the Keck Institute for Space Studies ``In Situ Science and Instrumentation          
for Primitive Bodies'' study group, including Bethany Ehlmann, Jordana Blacksberg, and John Eiler, were extremely valuable in the development of the model presented in this paper. We also thank Mario Melita for providing helpful comments and suggestions during the review process.

\end{document}